\documentclass[twocolumn,aps,showpacs]{revtex4}
\usepackage{graphicx}
\usepackage{dcolumn}
\usepackage{color}
\usepackage{amsmath}
\usepackage{bm}
\usepackage{amssymb}
\def\prl#1#2#3{{ Phys. Rev. Lett.} {\bf #1}, #2 (#3)}

\def\etl{$et ~al.$}

\def\bc{\begin{center}}
\def\ec{\end{center}}
\def\eqn{\end{equation}\noindent}
\def\eqnr{\end{eqnarray}\noindent}
\def\beqr{\begin{eqnarray}}
 
\begin{document}
\title{Driving--induced bistability in coupled chaotic attractors}
\author{Manish Agrawal$^{1}$, Awadhesh Prasad$^{2}$, and Ram Ramaswamy$^{3}$}
\affiliation{$^1$Department of Physics, Sri Aurobindo College, University of Delhi, New Delhi 110017, India\\
$^{2}$Department of Physics and Astrophysics, University of Delhi, Delhi 110007, India\\
$^{3}$University of Hyderabad, Hyderabad 500046, India}

\begin{abstract}
We examine the effects of symmetry--preserving and breaking interactions in a drive--response system where the response has an invariant symmetry in the absence of the drive. Subsequent to the onset of generalized synchronization, we find that there can be more than one stable attractor. Numerical, as well as  analytical  results establish the presence of phase synchrony in such coexisting attractors. These results are robust to external noise. 
\end{abstract}

\pacs{05.45.Ac, 05.45.Xt, 05.45.Pq}

\maketitle

\section{Introduction}
Synchronization is an ubiquitous phenomenon in the dynamics of interacting nonlinear systems. Although first used to denote the establishment of identical dynamics in weakly coupled systems, 
in recent decades it has been realized that the concept extends itself to describe correlated motion in coupled systems quite naturally.  Based on the nature of the systems and the manner in which they are coupled, several types of synchronization are known \cite {pecora, pikovsky1, general, fuzisaka, pikovsky2, rulkov, rosen1, garima}. 

Systems can be coupled either mutually or one--way as in the ``drive-response" scenario wherein the response is enslaved by the forcing (master) system. In this latter case, the equations of motion can usually
be represented in the following skew--product form,
\begin{eqnarray}
\label{eq:drivereponse}
\dot{\textbf{X}}&=&\textbf{F}(\textbf{X})\\
\ ~ \nonumber\\
\dot{\textbf{Y}}&=&\textbf{G}(\textbf{Y}, \textbf{X}),
\end{eqnarray}
where \textbf{X} and \textbf{Y} are the dynamical variables 
of the drive and the response systems respectively, and \textbf{F} and \textbf{G} are continuous and differentiable vector fields.  The occurrence of generalized synchronization (GS) in such systems implies the existence of a unique (but not necessarily differentiable) functional relationship \textbf{Y} = $\boldsymbol\Psi$(\textbf{X}) between the variables of the drive and response systems \cite {rulkov}.  
The onset of GS occurs when the largest conditional Lyapunov exponent (LCLE) of response system becomes negative \cite{kocarev, pyragas}.  The presence of GS is further determined by constructing an auxiliary copy ($\textbf{Y}^{\prime}$) of the response system as
\begin{eqnarray}
\label{eq:reponse}
\dot {\textbf{Y}^{\prime}}&=&\textbf{G}(\textbf{Y}^{\prime}, \textbf{X}),
\end{eqnarray} 
where the occurrence of GS between drive (\textbf{X}) and response (\textbf{Y}) is confirmed by observing complete synchronization (CS) between the response system (\textbf{Y}) and its auxiliary copy (\textbf{Y}$^{\prime}$) \cite{abarbanel}. 

When non--identical systems are coupled, they can exhibit a form of correlated dynamics that is termed phase synchronization (PS): the phases of oscillation can become entrained while the amplitudes remain uncorrelated \cite{rosen1}. This form of synchrony is quite common in natural systems wherein there is a range of time--scales in the dynamics.    

Studies of GS have largely focused on chaotic systems having single-scroll dynamics  where the oscillation is around a single fixed point. However systems with multi-scroll dynamics when the trajectory hops around more than one  fixed point are also known, and are also known to be shown by the important class of systems having inherent symmetries. Guan \etl \cite{guan} have considered such symmetric systems in drive-response coupling scenario and observed bistability within the GS regime. However, the dynamics of these bistable attractors has not been studied in detail. 
        
The motivation of the present work  is the study of induced bistability in unidirectionally coupled symmetric response systems. The coupling may either preserve or break the symmetry of the response system, but we find that the resulting bistable attractors are always phase-synchronized.
Our results further elucidate the nature of GS and extend the auxiliary system approach to the analysis of phase synchrony between response system and its auxiliary copy. These observations are 
seen to be robust to noise.

The paper is organized in following manner. The possible coupling schemes in context of symmetry of response system are discussed in Sec. II. This is followed by numerical and analytical demonstration of phase relations with different coupling schemes in Sec. III and IV. The robustness of these results is verified in Sec. V. The paper concludes with a summary in Sec. VI.

\section{coupling schemes}
In present work, we are interested in studying the response behavior of a particular class of systems, namely those which are invariant with respect to some symmetry transformations.  A familiar example is provided by the Lorenz oscillator  \cite{lorenz}, the governing equations of which are
\begin{eqnarray}
\dot {x}&=&\sigma(y-x)\nonumber \\
\dot {y}&=&rx-y-xz\nonumber \\
\dot {z}&=&xy-\beta z.
\label{eq:lor}
\end{eqnarray} \noindent
This system is invariant under the transformation $T:(-x,-y,z)$ $\rightarrow$ $(x,y,z)$, and thus the attractor is symmetric about the $z$- axis, or equivalently it has inversion symmetry in the $x-y$ plane.
As is well known, this system has a ``butterfly'' shaped strange attractor (at parameters $\sigma$=$10$, $r$=$28$, and $\beta$=$8/3$), where the trajectory moves between quadrants 1 and 3 in $x-y$ plane.

We consider driving this system by an external dynamical system, with diffusive scalar coupling \cite{heagy}, where the drive is coupled to a single variable of response system.  Clearly  the response system can be forced in a manner that either preserves or breaks the symmetry. In the present case, if the  (unidirectional) coupling is introduced in the $z$ variable then the symmetry in $x$ and $y$ is preserved, and this coupling  is termed \textit{symmetry preserving}.  If the response is coupled through either the $x$ or the $y$ variables, the inversion symmetry with respect to the $z$ axis or $x-y$ plane is lost and this is a \textit{symmetry breaking} coupling. The dynamical behavior of response system under these two coupling schemes are studied in next sections. 
 
\section{Symmetry--preserving interaction}

Consider the Lorenz system  (Eq. \ref{eq:lor}) driven by a chaotic R\"ossler \cite{rossler} oscillator, the complete system being given by
\begin{eqnarray}
\label{eq:ross}
\dot {x}{_1}&=&-y_1-z_1\nonumber \\
\dot {y}{_1}&=&x_1+ay_1 \nonumber \\
\dot {z}{_1}&=&b+z_1(x_1-c)\nonumber \\
\ ~ \nonumber\\
\dot {x}{_2}&=&\sigma(y_2-x_2)\nonumber \\
\dot {y}{_2}&=&rx_2-y_2-x_2z_2\nonumber \\
\dot {z}{_2}&=&x_2y_2-\beta z_2+\epsilon(z_1-z_2).
\end{eqnarray}
\noindent

\begin{figure}
\includegraphics [scale=0.32]{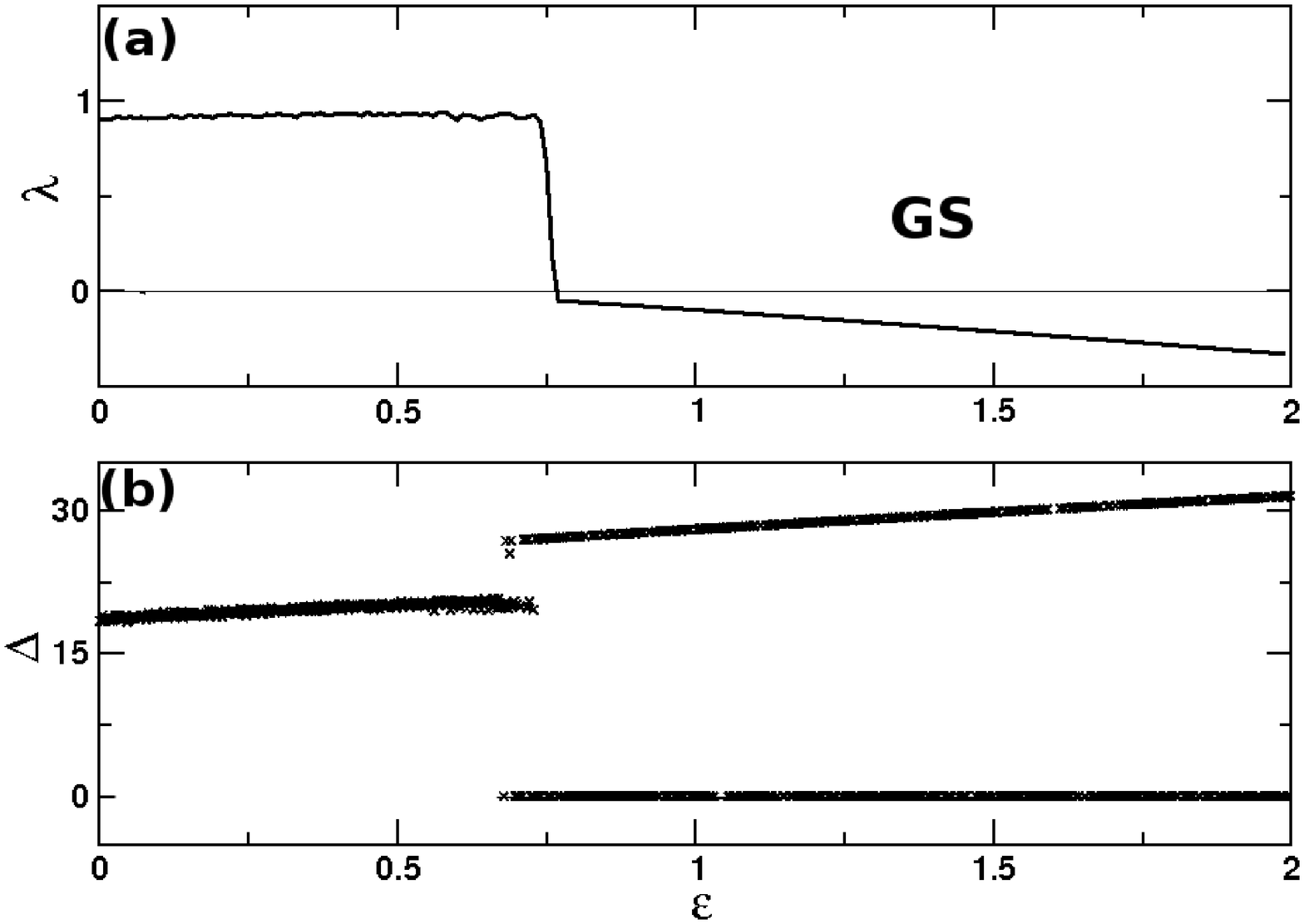}
\caption{Lorenz (response) system: (a) the largest conditional Lyapunov exponent ($LCLE$) and (b) the average synchronization error ($\Delta$) between response system and its auxiliary copy, with coupling parameter $\epsilon$.}  
\label{fig:lorenz1}
\end{figure}

\begin{figure}
\includegraphics [scale=0.3]{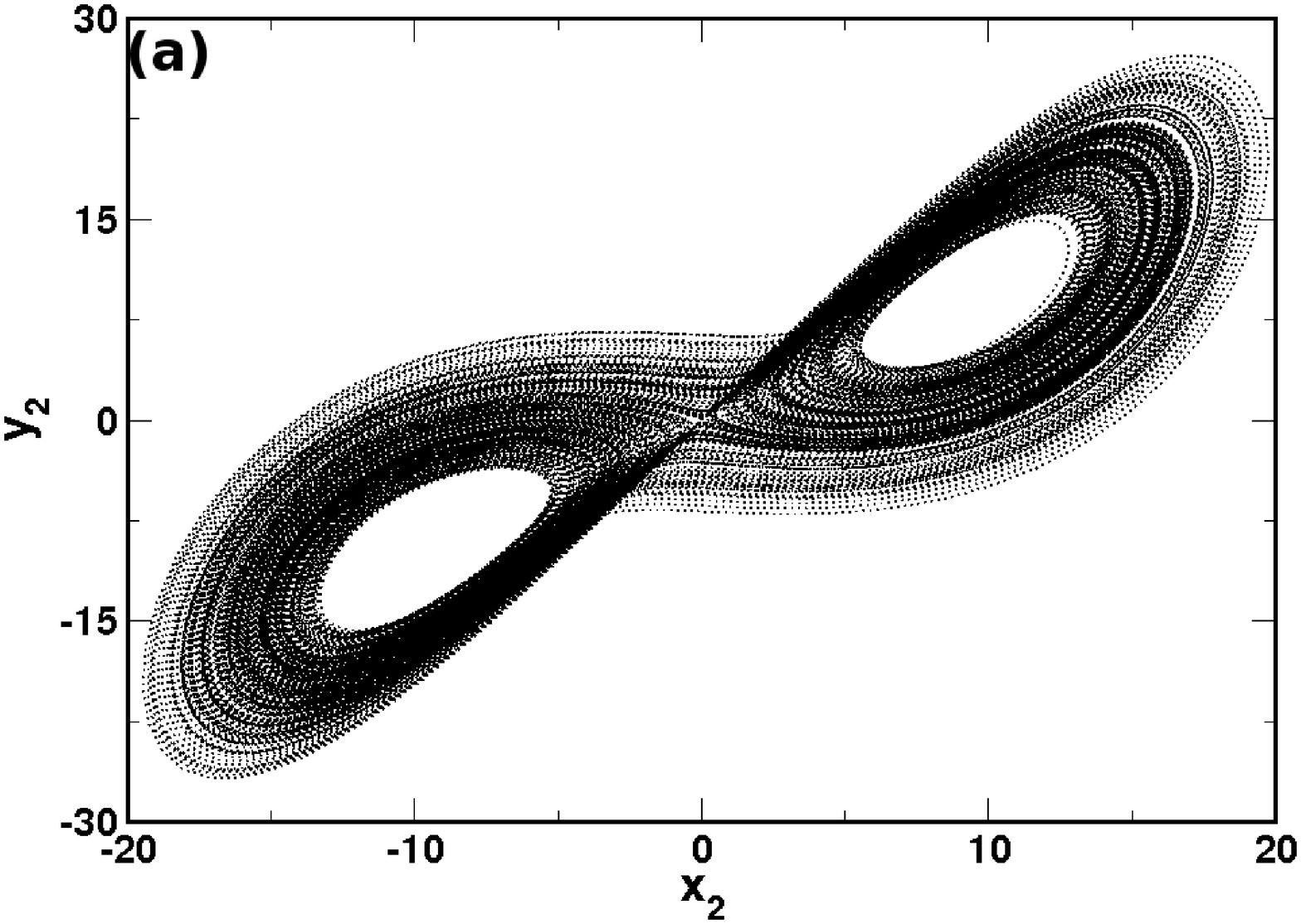}
\includegraphics [scale=0.3]{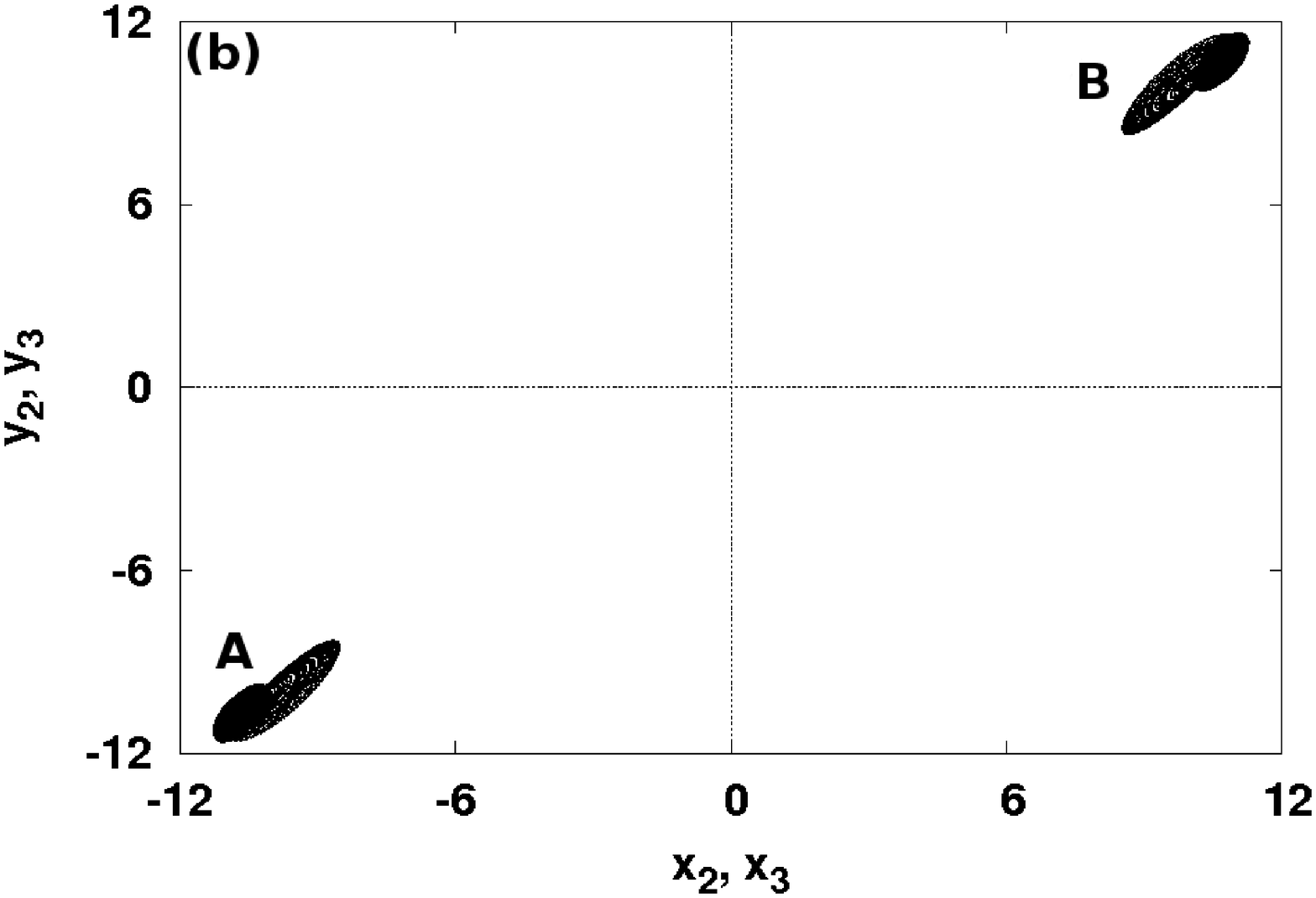}
\includegraphics [scale=0.3]{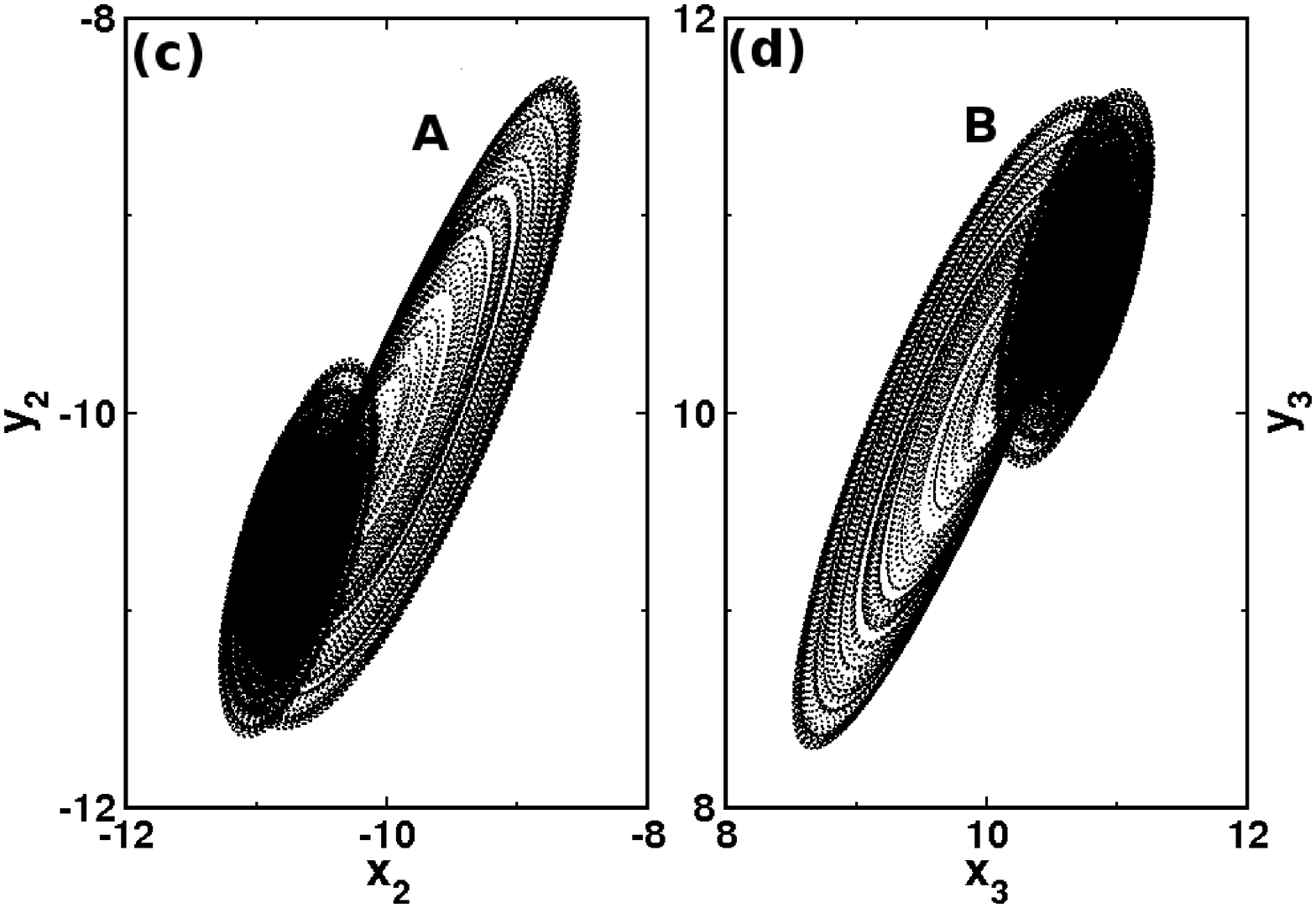}
\caption{(a) Lorenz attractor before the onset of GS at coupling strength $\epsilon$=$0.74$, (b) The symmetric pair of coexisting attractors $A$ and $B$ in response system for model Eqs.(\ref{eq:ross}), (c) , and (d) the expanded views of coexisting attractors $A$ and $B$. Coupling parameter $\epsilon$=$1.59$.}  
\label{fig:traj}
\end{figure}

The symmetry of the response Lorenz system in the $x_2-y_2$ plane is preserved by the coupling. 
Fixing the internal parameters of drive and response oscillators to standard values, $a$=0.2, $b$=0.2, and $c$=5.7 and $\sigma$=10, $r$=28, and $\beta$=8/3 respectively, both systems have chaotic dynamics. The largest conditional Lyapunov exponent (LCLE)\cite{kocarev, pyragas} as a function of the  coupling parameter $\epsilon$ is shown in Fig. \ref{fig:lorenz1}(a). 
With the increase in coupling parameter the LCLE becomes negative at $\epsilon$ $\approx$ 0.767, which indicates the onset of  GS between the drive and response systems.  By constructing an auxiliary system \cite{abarbanel} and measuring the average synchronization error ($\Delta$) between the response system and its auxiliary copy confirms GS. The error $\Delta$ \cite{yang} is defined as 
\begin{eqnarray}
\label{eq:ase1}
\Delta =\langle\sqrt{((x_2-x_3)^{2}+(y_2-y_3)^{2}+(z_2-z_3)^{2})}\rangle,\nonumber
\end{eqnarray}
\noindent 
where $\langle\cdot\rangle$ denotes a time average as well as over an ensemble of initial conditions, typically taken to be 100.  The value of $\Delta$ as a function of $\epsilon$  are  shown in Fig. \ref{fig:lorenz1}(b), and at the transition to GS, $\Delta$ takes two values, indicating that there are two distinct attractors, one on which  $\Delta=0$ and one on which $\Delta>0$. The trajectories corresponding to the different attractors are shown in  Figs. \ref{fig:traj}(a) and (b),  for  $\epsilon$=0.74 and 1.59 respectively. 

Figure \ref{fig:traj}(a) shows that there is always  a single attractor irrespective of initial conditions, prior to the onset of GS, but as can be seen in Fig. \ref{fig:traj}(b), the response system consists
of two \textit{symmetric} coexisting attractors $A$ and $B$ that are spatially separated. 
Symmetry preservation is clearly shown in Figs. \ref{fig:traj}(c) and (d) corresponding to attractors $A$ and $B$ respectively. With different initial conditions, if both the response system and its copy evolve to either the attractor $A$ or  attractor $B$ then $\Delta = 0$.
However it is also possible that the trajectories of the response  and the auxiliary converge to 
different attractors, and this gives a positive average synchronization error, namely $\Delta > 0$, even though there is GS.

\begin{figure}
\includegraphics [scale=0.3]{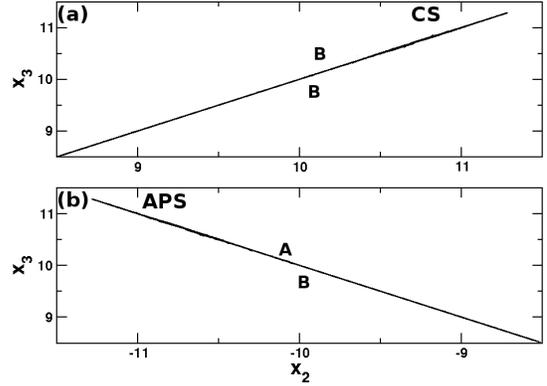}
\caption{Lorenz attractor: the evolution of response system ($x$ variable) for two different sets of initial conditions at $\epsilon$=$1.59$, where these conditions converges to (a) the same attractor $B$ ($CS$, $x_2=x_3$), (b) either of symmetric coexisting attractors $A$ and $B$ ($APS$, $x_2=-x_3$). The parameters are that of the Fig. 2(b).}  
\label{fig:tseries}
\end{figure}

This is termed as symmetry induced bistability, and the occurrence of two coexisting attractors $A$ and $B$ can be verified for a number of different initial conditions. One question of interest is the phase relationship between the distinct attractors  (when $\Delta>0$). Shown in Figs. \ref{fig:tseries}(a) and (b) are the trajectories in $x_2-x_3$ plane (to infer the relative phase) when both the response system and its copy, lead to the same attractor (here $A$)  or to different attractors ($A$ and $B$) 
respectively for two sets of different initial conditions. Fig. \ref{fig:tseries}(a) shows that the response system and its auxiliary copy are in-phase or in complete synchrony ($x_2=x_3, y_2=y_3$) which gives $\Delta=0$,  while, Fig. \ref{fig:tseries}(b) presents the case when different initial conditions settle onto coexisting attractors ($A$ and $B$). The response and the auxiliary are in anti--phase synchrony (APS) since $x_2=-x_3$ and $y_2=-y_3$ . 

A phase variable can be defined for the the Lorenz system as 
$\phi(t)=\arctan (z-z_{0})/(u-u_{0}) $ by considering the projection of trajectory
onto the $(u,z)$ plane where $u =\sqrt{x^{2}+y^{2}}$ and ($u_{0}, z_{0}$) is the enclosed fixed point \cite{pikovsky}. We denote the phases of the response and auxiliary as $\phi_{2}$ and $\phi_{3}$  respectively. 
To establish the phase relationships between the response and its copy we begin with the trigonometric identity,
\begin{eqnarray}
\label{eq:function2}
 \tan (\phi_{2}-\phi_{3})=\dfrac{\tan \phi_{2}-\tan \phi_{3}}{1+ \tan \phi_{2} \tan \phi_{3}},
\end{eqnarray}
\noindent
using which one can write the phase relation as
\begin{eqnarray}
\label{eq:function3}
\tan (\phi_{2}-\phi_{3})=\dfrac{\dfrac{z_{2}-{z_0}_2}{u_2-{u_0}_2}-\dfrac{z_3-{z_0}_3}{u_3-{u_0}_3}}{1+\dfrac{z_2-{z_0}_2}{u_2-{u_0}_2}                                 \dfrac{z_3-{z_0}_3}{u_3-{u_0}_3}}\nonumber\\
=\dfrac{(u_3-{u_0}_3)(z_2-{z_0}_2)-(u_2-{u_0}_2)(z_3-{z_0}_3)}{(u_2-{u_0}_2)(u_3-{u_0}_3)+(z_2-{z_0}_2)(z_3-{z_0}_3)}.
\end{eqnarray}
\noindent
For any $\Delta \ge $ 0,  when $\phi_{2}-\phi_{3}$ 
is 0 or $\pi$,  $\tan (\phi_{2}-\phi_{3})$=0, giving  
\begin{eqnarray}
\label{eq:function4}
(u_3-{u_0}_3)(z_2-{z_0}_2)-(u_2-{u_0}_2)(z_3-{z_0}_3)=0.
\end{eqnarray}
\noindent

Since the Lorenz attractor has the inversion symmetry $T:(-x,-y,z)$ $\rightarrow$ $(x,y,z)$ in $x-y$ plane,  
the $z$ variables of the response system and the auxiliary  remain identical in GS regime.
Setting $z_3=z_2$ and ${z_0}_3={z_0}_2$ Eq. (\ref{eq:function4}) can be rewritten  as 
\begin{eqnarray}
\label{eq:function5}
(z_2-{z_0}_2)[(u_3-{u_0}_3)-(u_2-{u_0}_2)]=0.
\end{eqnarray}
\noindent     
Here, $z_2\neq {z_0}_2$ because ${z_0}_2$ is the $z-$ component of the unstable fixed point.
It is easy to show that this leads to the condition  
\begin{eqnarray}
\label{eq:function9}
(x_3+x_2)(x_3-x_2)+(y_3+y_2)(y_3-y_2)=0,
\end{eqnarray}  
\noindent 
which can be satisfied for both CS, namely $\phi_{2}-\phi_{3}=0$ when $x_3$=$x_2$ and $y_3$=$y_2$  as well as for  $x_3=-x_2$ and $y_3=-y_2$ namely APS. Both cases are realized by our numerical results. 

\section{Symmetry--breaking interaction}

Consider the coupling in $x$  in the response, namely the equations of motion 
\begin{eqnarray}
\label{eq:xcouple}
\dot {x}{_2}&=&\sigma(y_2-x_2)+\epsilon(x_1-x_2)\nonumber \\
\dot {y}{_2}&=&rx_2-y_2-x_2z_2\nonumber \\
\dot {z}{_2}&=&x_2y_2-\beta z_2.
\end{eqnarray}

\begin{figure}
\includegraphics [scale=0.32]{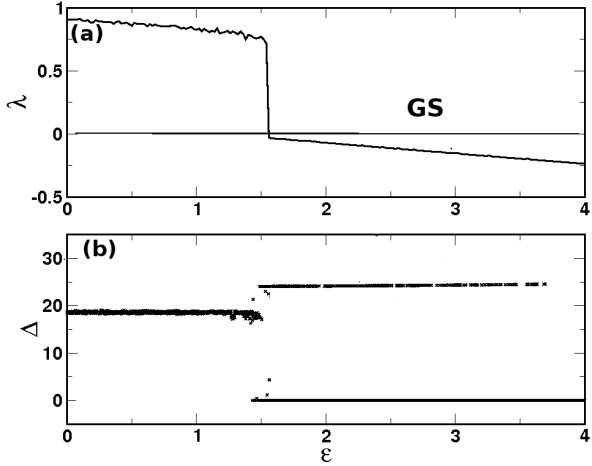}
\caption{Lorenz (response) system in symmetry breaking case: (a) the largest conditional Lyapunov exponent ($LCLE$) and (b) average synchronization error ($\Delta$), with coupling parameter $\epsilon$.}  
\label{fig:lorenz4}
\end{figure}

\begin{figure}
\includegraphics [scale=0.35]{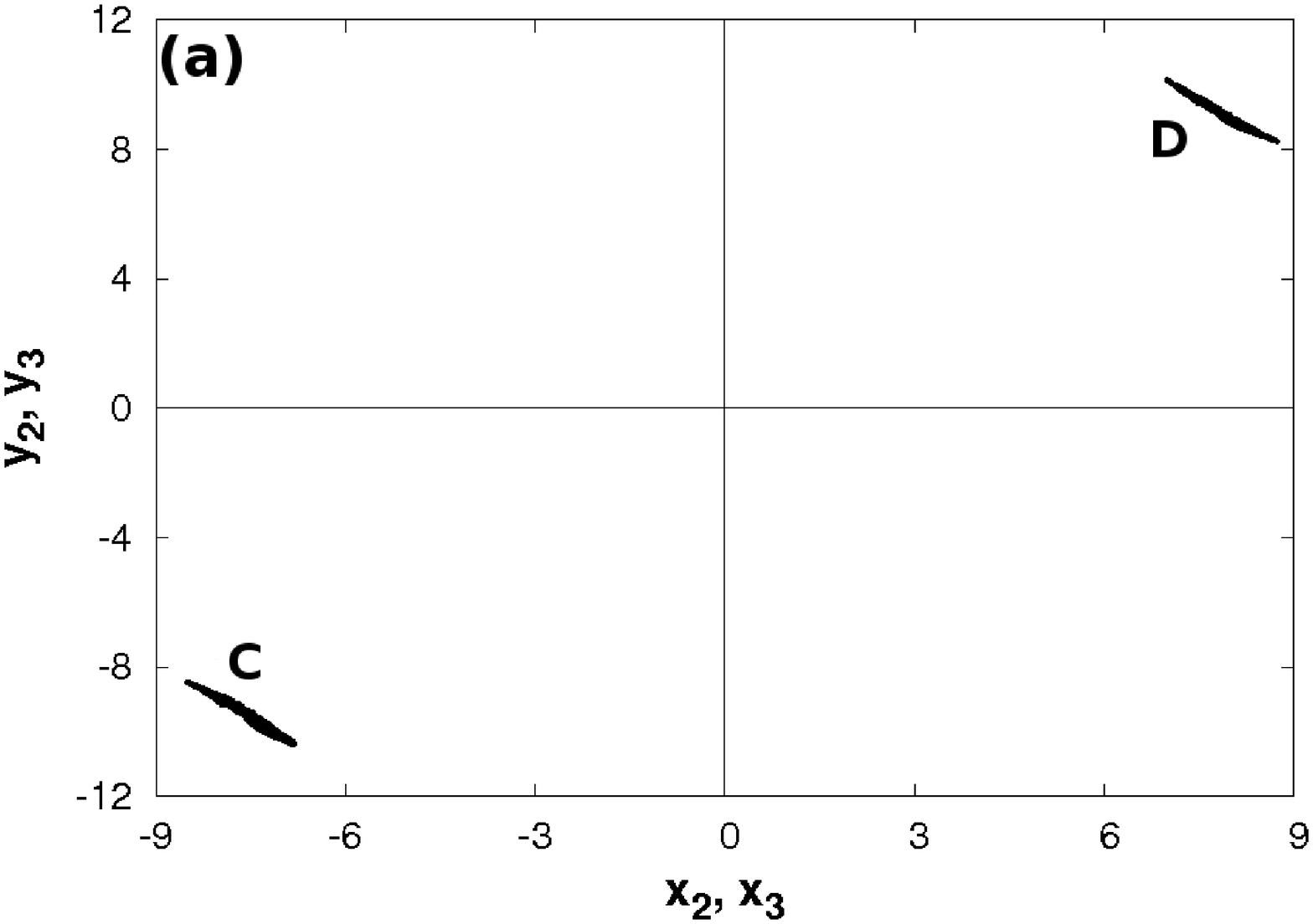}
\includegraphics [scale=0.32]{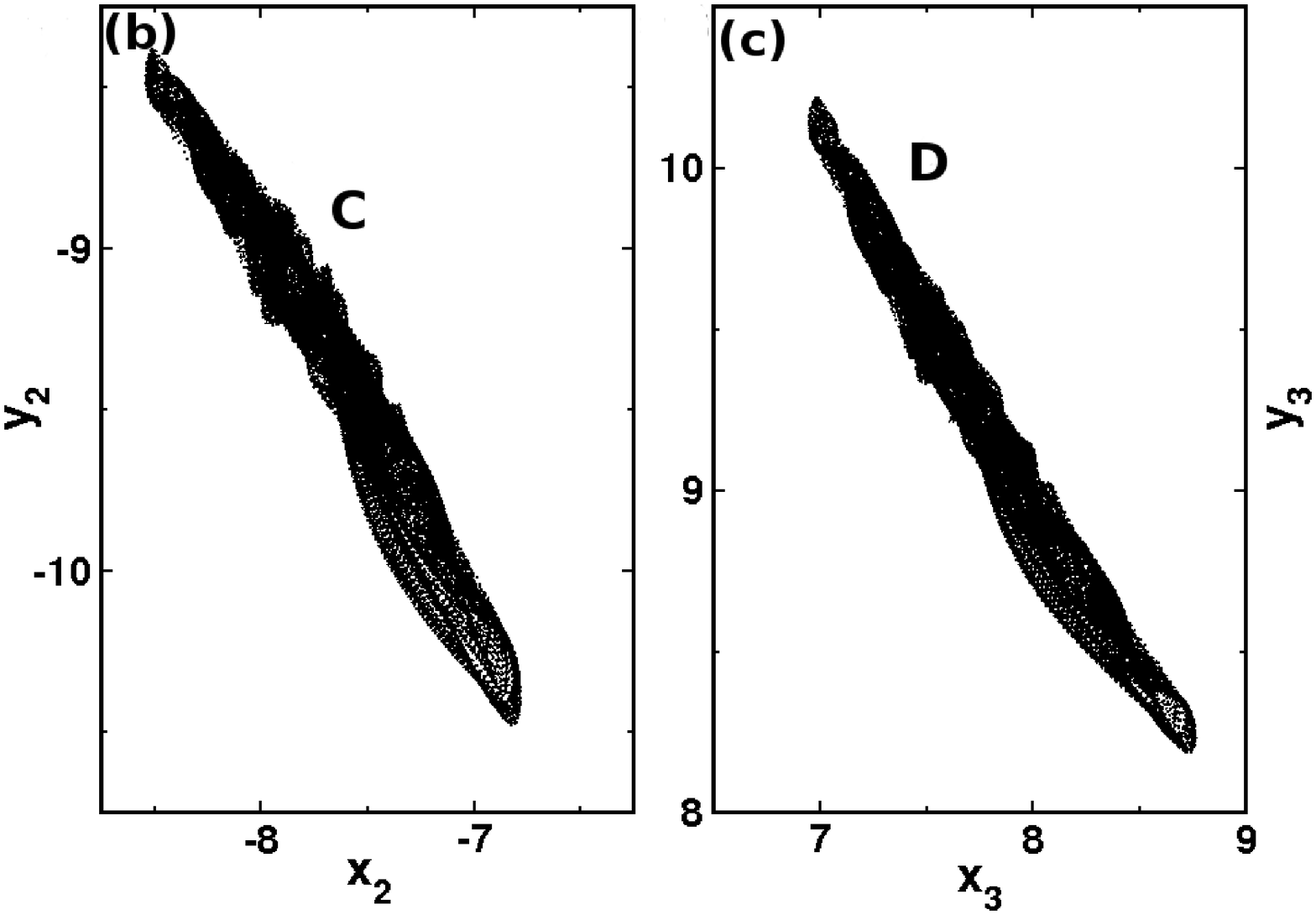}
\caption{(a) Asymmetric pair of coexisting attractors $C$ and $D$ in ($x$- coupled) symmetry breaking case, (b) and (c) the expanded views of coexisting attractors $C$ and $D$. Coupling parameter $\epsilon=2.0$.}  
\label{fig:asym1}
\end{figure}

The internal parameters of drive and response systems are kept the same as in Eq. (\ref{eq:ross}).
Figs.~\ref{fig:lorenz4}(a) and \ref{fig:lorenz4}(b) show the LCLE and $\Delta$ respectively for
 this coupling scheme. As discussed in earlier section, the fluctuation of average synchronization 
error ($\Delta$) having the values  $\Delta > 0$ implies the presence of bistable chaotic attractors. As shown in Fig. \ref{fig:asym1}(a), the response system consists of two coexisting \textit{asymmetric} attractors $C$ and $D$ (in GS regime) for two sets of different initial conditions. 
However, these attractors, $C$ and $D$, are asymmetric in phase space as clearly shown in
expanded Figs. \ref{fig:asym1}(b) and (c) respectively.

\begin{figure}
\includegraphics [scale=0.3]{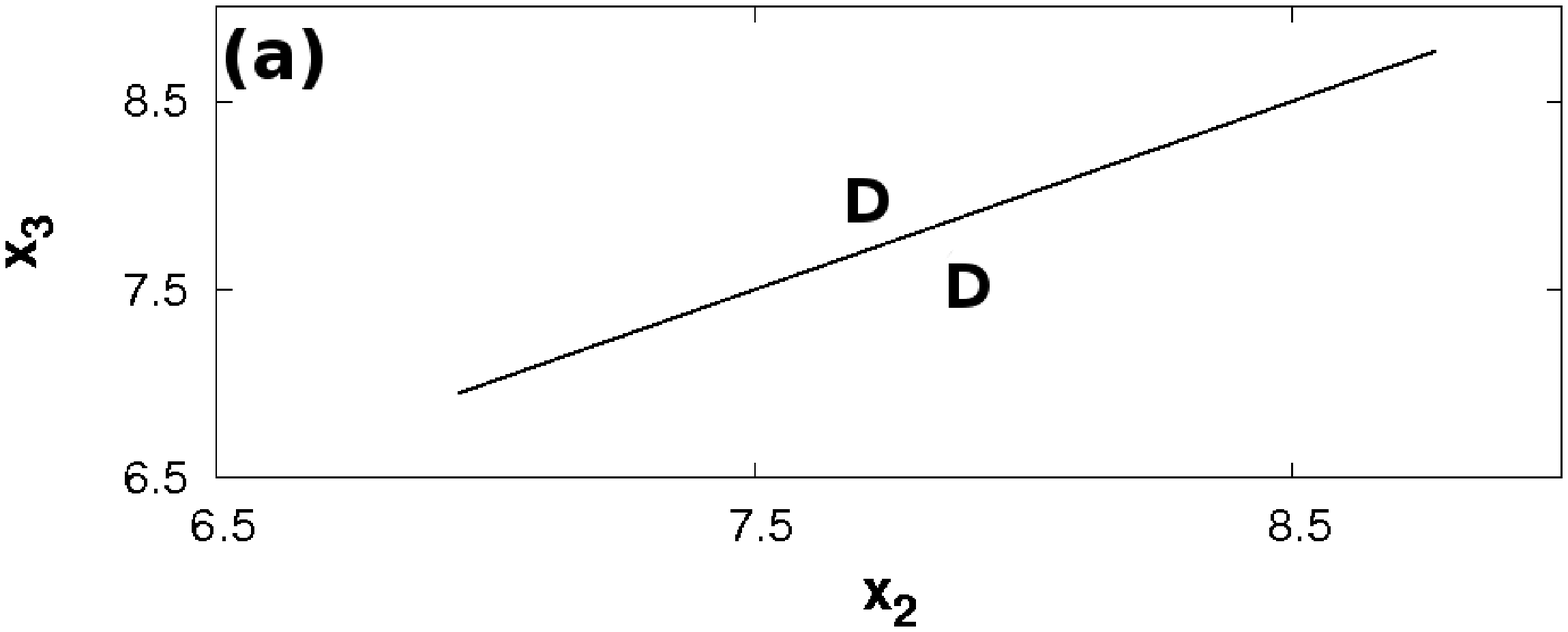}
\includegraphics [scale=0.3]{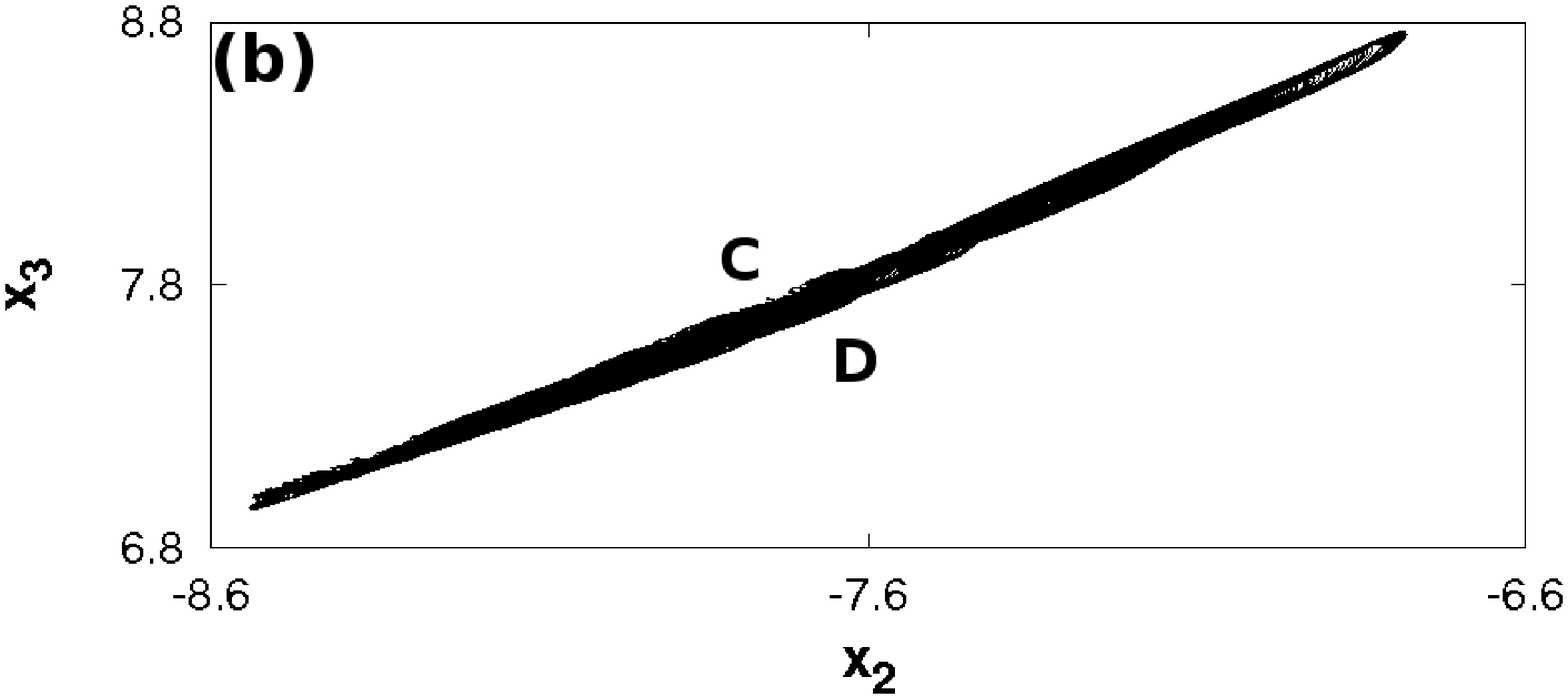}
\caption{Symmetry breaking interaction: the evolution of response system ($x$ variable) for two different initial conditions, where these conditions converge to (a) the same attractor $C$ ($\Delta=0$), (b) either of asymmetric coexisting attractors $C$ and $D$ ($\Delta>0)$). Coupling parameter $\epsilon=2.0$.}  
\label{fig:asym2}
\end{figure}

The synchrony properties of the asymmetric coexisting attractors $C$ and $D$ 
when $\Delta > 0$ in the GS regime are of interest.      
Fig.~\ref{fig:asym2}(a) is for the case when both response and auxiliary lead to the same attractor (denoted $D$) for two sets of different initial conditions: the response  and auxiliary  are {\it in-phase} or in complete synchrony ($x_2=x_3, y_2=y_3$). Fig.~\ref{fig:asym2}(b) is for the case when the response and  auxiliary converge to distinct attractors: the two systems are in-phase. 
\section{Persistence with Noise}
In order to see the constancy of results obtained in case of symmetry preserving and breaking interactions, the effects of added noise are considered. 
Thus in the symmetry preserving case (cf. Eq. (\ref{eq:ross})) we consider the system 
\begin{eqnarray}
\label{eq:znoisecouple}
\dot {x}{_2}&=&\sigma(y_2-x_2)+\xi S_1\nonumber \\
\dot {y}{_2}&=&rx_2-y_2-x_2z_2+\xi S_2\nonumber \\
\dot {z}{_2}&=&x_2y_2-\beta z_2+\epsilon(z_1-z_2)+\xi S_3,
\end{eqnarray}
and for symmetry breaking case (cf. Eq. (\ref{eq:xcouple})),
\begin{eqnarray}
\label{eq:xnoisecouple}
\dot {x}{_2}&=&\sigma(y_2-x_2)+\epsilon(x_1-x_2)+\xi S_4\nonumber \\
\dot {y}{_2}&=&rx_2-y_2-x_2z_2+\xi S_5\nonumber \\
\dot {z}{_2}&=&x_2y_2-\beta z_2+\xi S_6.
\end{eqnarray} 
Here $S_i$ ($i=1,2,..,6$) is $\delta$--correlated random noise of strength  $\xi$.  Other parameters ($\sigma$, $r$, $\beta$) are of the same values as used earlier in Sections II and III. 
\begin{figure}
\includegraphics [scale=0.35]{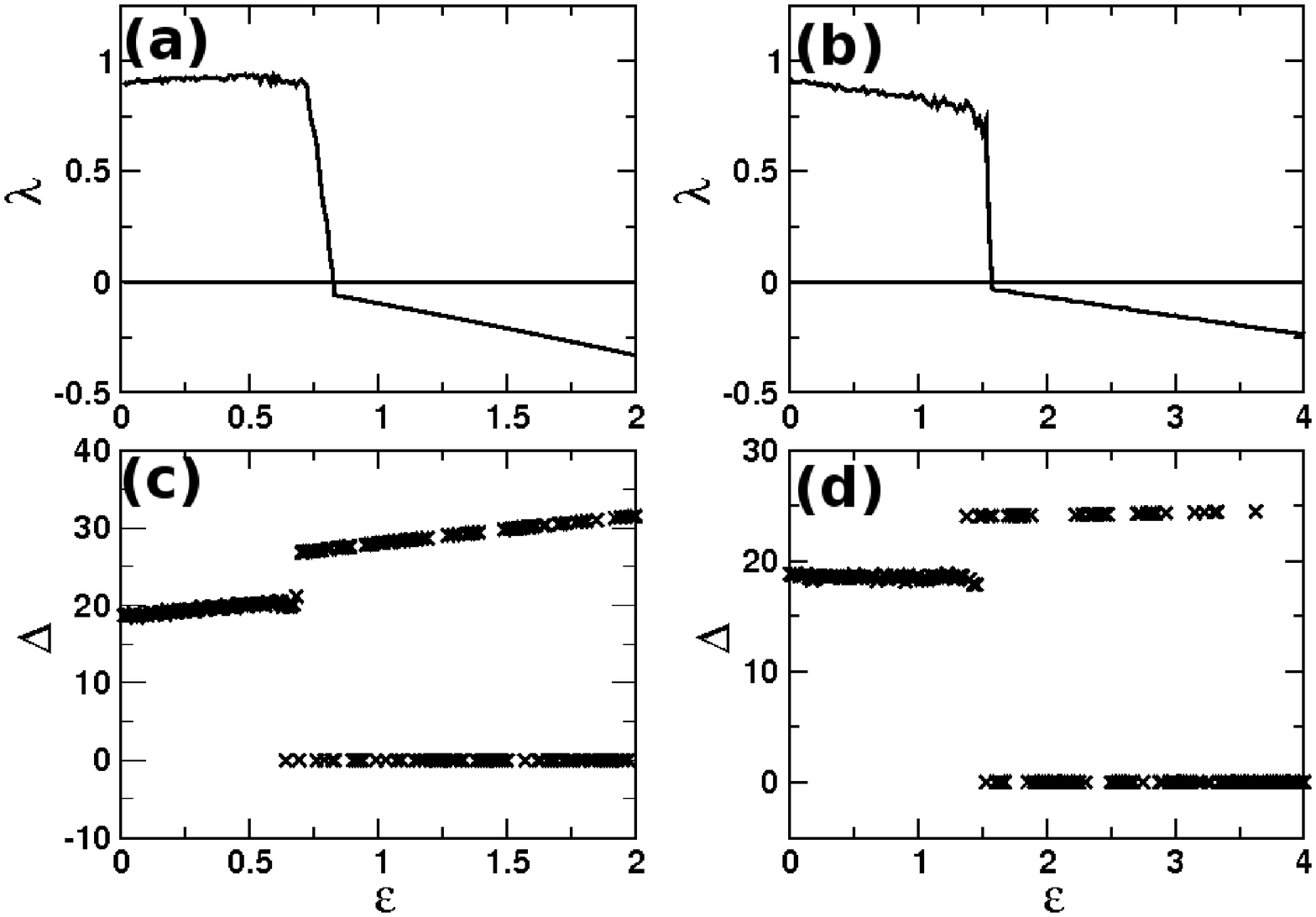}
\includegraphics [scale=0.35]{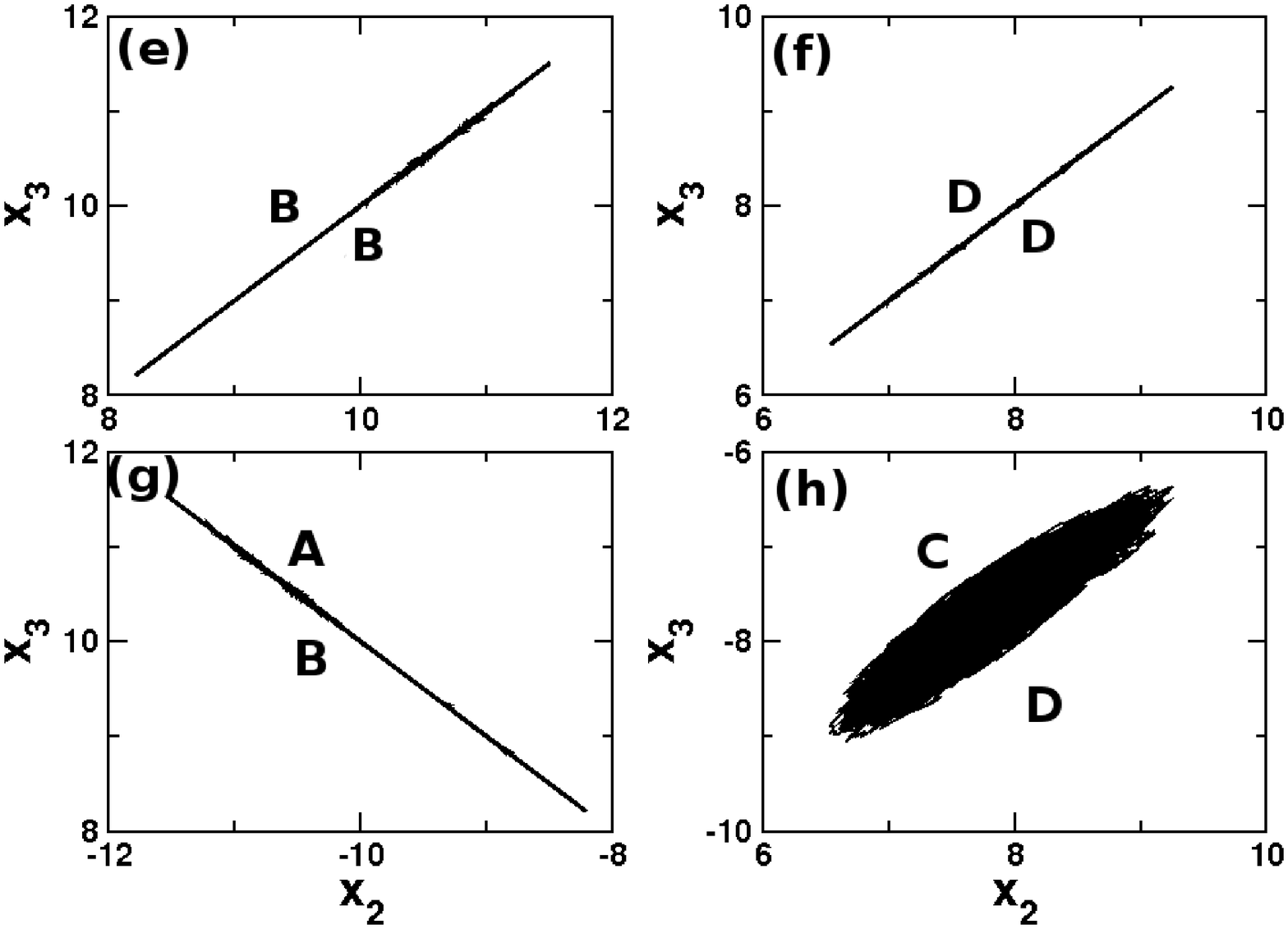}
\caption{ The left and right panels correspond to the symmetry preserving and symmetry breaking interaction respectively
with noise ($\xi=10.0$). (a) and (b) show the largest conditional Lyapunov exponents ($LCLE$) while
(c) and (d) show the average synchronization error ($\Delta$) as a function of coupling strength, $\epsilon$.
(e) and (f): the trajectories in ($x_2$, $x_3$) plane, when the response and its auxiliary copy  
converge to the same attractors $B$ and $D$ respectively.
(g) and (h): the trajectories in ($x_2$, $x_3$) plane, when the response and its auxiliary copy 
converge to the  different  attractors  $A$ or $B$ and $C$ or $D$ respectively. The coupling
parameter in (e-h) is $\epsilon=2$.}
\label{fig:spbnoise}
\end{figure} 
The effects of noise for symmetry preserving and symmetry breaking interactions are illustrated 
in left and right panels in Fig. \ref{fig:spbnoise} respectively with noise strength, $\xi =10.0$.
The variation of LCLE with coupling parameter, $\epsilon$ are shown in Figs. \ref{fig:spbnoise} (a) and (b) which clearly
show that there is no significant difference as compared to Figs. \ref{fig:lorenz1} (a) and \ref{fig:lorenz4} (a) respectively.
The fluctuations of average synchronization error, $\Delta$ (Figs. \ref{fig:spbnoise} (c) and (d)) infer the occurrence of driving induced bistability even in presence of noise.

The phase relations between coexisting attractors are shown in Figs. \ref{fig:spbnoise} (e)--(h).
Figs. \ref{fig:spbnoise} (e) and (f) show the PS state in presence of noise, similar to Figs. \ref{fig:tseries} (a) and \ref{fig:asym2} (a) respectively. Figure \ref{fig:spbnoise} (g)
indicates the APS state similar to Fig. \ref{fig:tseries} (b), while Fig.\ref{fig:spbnoise} (h) shows the PS state as in Fig. \ref{fig:asym2} (b).
We have also obtained the similar results even with higher noise strength.
Thus the results obtained in section II and III are robust to added noise, and hence implying the possibility of theirs experimental verification.   

\section{Summary}
In present work we have studied the effect of external forcing on systems with invariant symmetry, using the Lorenz equations as a prototypical example. The coupling has been introduced in two different ways, to keep the inherent symmetry of the driven system, or to break the symmetry. 

The onset of generalized synchronization between the drive and the response is studied,  and we find that there is bistability {\it after} GS has set in: this is clearly drive induced. The coexisting attractors have been found to be symmetric or asymmetric depending upon the coupling, and we use the auxiliary system approach to verify our results.  In the symmetry preserving case, we find that the attractors in the response and the auxiliary can be either in phase \textit{or} antiphase, and we are able to present analytical arguments for these cases. However for the symmetry breaking interaction the coexisting attractors (which are asymmetric) can \textit{only} have in-phase synchrony.  

In this paper we have only used the chaotic R\"ossler attractor as a drive, but it should be noted that   results are similar  for any drive that has suitable internal timescales. Further, although, we have taken the response to be the Lorenz system here, we have observed similar results  for phase synchronization states in other  inversion symmetric response systems \cite{magrawal}, and we believe that our results will hold quite generally. 
                 
\begin{acknowledgments}

MA is grateful to University Grants Commission (UGC, India), and AP \& RR would like to thank the DST, Govt. of India for financial support. AP also acknowledges the financial support from the DU-DST PURSE grant.

\end{acknowledgments}


\begin{thebibliography}{99}
\bibitem{magrawal}
We have considered other systems such as the two--scroll, inversion--symmetric
model studied by Chen \cite{chen} or the Rucklidge system \cite{rucklidge}, or
for that matter the
single-scroll Sprott  \cite{sprott} as the response and the R\"ossler oscillator
as a drive. Similar behavior was seen, namely in--phase and anti-phase
synchronized dynamics for the symmetric preserving situation  and only
in--phase motion for the symmetry--breaking case.
\end{thebibliography}
\end{document}